\documentclass{article}

\usepackage[dblblindworkshop, final]{neurips_2026}

\usepackage[utf8]{inputenc} %
\usepackage[T1]{fontenc}    %
\usepackage{hyperref}       %
\usepackage{url}            %
\usepackage{booktabs}       %
\usepackage{amsfonts}       %
\usepackage{amsmath}        %
\usepackage{nicefrac}       %
\usepackage{microtype}      %
\usepackage{xcolor}         %
\usepackage{multirow}       %
\usepackage{graphicx}
\usepackage{tcolorbox}
\title{Aggregate Accuracy Hides Diagnosis-Specific Failure in Medical Vision--Language Models}

\workshoptitle{Medical Reasoning with Vision Language Foundation Models (Med-Reasoner)}

\author{%
  Shreya Ajay Kale\thanks{Equal contribution; joint first authors.} \\
  Independent Researcher \\
  \texttt{129shreya@gmail.com} \\
  \And
  Angad Ripudaman Singh Bajwa\footnotemark[1] \\
  Independent Researcher \\
  \texttt{angadisthere@gmail.com} \\
}

\begin{document}

\maketitle

\begin{abstract}
Medical vision--language models (VLMs) are compared mainly by accuracy averaged across findings, a summary that can hide complete failure on a single diagnosis. We audit five open medical VLMs (MedGemma-4B, Lingshu-7B, HuatuoGPT-Vision-7B, Qwen2.5-VL-7B, and MedVLM-R1) on 6{,}048 ChestX-ray14 questions balanced by sex, age, finding, and label. On the 432 pneumothorax-positive X-rays, HuatuoGPT-Vision and Qwen2.5-VL detect zero cases while MedGemma and Lingshu detect 38\% and 67\% of the same images. The two failing models answer ``no'' to all 864 pneumothorax cases (balanced accuracy 0.500, chance), yet produce fluent, anatomically specific explanations that name the signs of a collapsed lung and then deny them, with no hedge or abstention. This failure is not an artifact of prompting, image resolution, label ambiguity, or model scale. It persists across four prompt formats, at native resolution, and on single-label cases. A within-model control shows the same forced prompt gives HuatuoGPT 79\% recall on nodule but only 6\% on pneumothorax. It survives a tenfold scale increase (Qwen2.5-VL-72B detects 4\%) and replicates on SIIM-ACR, an independent dataset with radiologist-drawn masks, where the failing models detect 0 of 432 confirmed pneumothoraces, including 0 of the 144 largest lesions. Reading the models' logits and internal states shows this is not blindness. The two failing models are highly confident yet no better than chance and a linear probe recovers pneumothorax from their hidden representations even where recall is zero, so the finding is represented internally but suppressed at the output. Aggregate accuracy therefore cannot establish that a medical VLM can see a given finding, because a model that misses one can still describe it convincingly. Reporting per-finding recall and specificity on balanced cases is needed to tell the difference.
\end{abstract}

\section{Introduction}
Medical vision--language models (VLMs) are increasingly proposed to read chest radiographs and are compared mainly through accuracy averaged over many findings. A single headline number is convenient and is what most readers use to judge a model.

An averaged number can hide complete failure on one finding, because averaging weights clinically different errors equally and a model can reach strong overall accuracy while never detecting one diagnosis. The failure is also hard to see in the output: a model can describe a radiographic sign of pneumothorax, such as the pleural line at the edge of a collapsed lung, and then state that it is absent, without refusing or hedging, so the response reads as a confident negative.

This matters because chest radiograph reading is high-stakes and these models are being positioned for triage. An untreated pneumothorax can be fatal within hours, yet a clinician reading a fluent, medically specific negative cannot distinguish it from a correct one, so the error is both consequential and hard to catch at the point of care.

Existing work has identified related problems but not this one. Hidden stratification shows that good average performance can hide failures on visual subtypes within a finding~\citep{oakdenrayner2020hidden}, and fairness audits report demographic performance gaps~\citep{seyyedkalantari2021underdiagnosis}. Neither asks whether a model with competitive overall accuracy can miss an entire finding while explaining it fluently, and medical VLM benchmarks still report aggregate accuracy without the per-finding check that would surface such a collapse. That averages can mask weak classes is itself well known. What is new here is a whole-diagnosis collapse delivered in fluent, confident language, together with a mechanism for it. Whether such a failure is only an artifact of prompting, image resolution, label noise, or model scale has also gone untested.

We audit five open medical VLMs, MedGemma-4B, Lingshu-7B, HuatuoGPT-Vision-7B, Qwen2.5-VL-7B, and MedVLM-R1, on a prevalence-matched set of 6{,}048 ChestX-ray14~\citep{wang2017chestxray} questions, and make four contributions. First, a prevalence-matched audit design that equalizes case mix across demographic groups. Second, evidence that aggregate accuracy hides finding-specific failure: MedGemma, the most accurate model overall, detects 95\% of atelectasis cases while incorrectly flagging 68\% of negatives, and two models detect no pneumothorax at all. Third, a prompt-invariant pneumothorax failure that replicates on an independent dataset with radiologist-drawn masks and survives a tenfold scale increase. Fourth, a mechanistic account of that failure: the failing models are confidently wrong rather than unsure, yet a linear probe decodes pneumothorax from their internal representations, locating the failure partly in a weaker representation and partly in output suppression.

\section{Related work}
Public chest X-ray datasets such as ChestX-ray14~\citep{wang2017chestxray} enabled large-scale training and benchmarking, but their labels were extracted from radiology reports rather than verified image by image. \citet{oakdenrayner2020exploring} documented label and content problems, including positive labels on post-treatment images, which affects how absolute recall should be interpreted. We therefore replicate our main result on SIIM-ACR Pneumothorax, in which a radiologist has segmented each pneumothorax at the pixel level. Work on hidden stratification~\citep{oakdenrayner2020hidden} similarly showed that good average performance can hide failures on clinically important subsets. Our finding-specific collapse reflects this problem at the level of the diagnostic finding rather than a visual subtype.

Fairness audits of chest X-ray models have reported demographic performance gaps~\citep{seyyedkalantari2021underdiagnosis}, but an apparent gap may reflect case mix rather than model behavior if groups differ in finding and label distributions. We address this by sampling the same finding-by-label grid for every demographic group.

The models evaluated here include both general-purpose and medically adapted VLMs. Qwen2.5-VL~\citep{qwen25vl2025} is a general-purpose model and also the backbone for Lingshu~\citep{lingshu2025} and for the HuatuoGPT-Vision~\citep{huatuogptvision2024} checkpoint we evaluate. MedGemma~\citep{medgemma2025} is built on a different base, and MedVLM-R1~\citep{medvlmr1_2025} is trained with reinforcement learning for medical reasoning.

\section{Method}

\subsection{ChestX-ray14 and prevalence-matched item construction}

We use ChestX-ray14~\citep{wang2017chestxray}, containing 112{,}120 frontal X-rays from 30{,}805 patients with finding labels, patient ID, age, sex, and view position. We removed 16 rows with implausible ages and retained AP and PA views. Labels follow the dataset annotations, with multi-label images positive for each labeled finding.

We crossed 2 sexes, 3 age bands (0--40, 40--60, and 60+), 7 findings, and 2 labels (present or absent) to form 84 cells, sampling 72 images per cell. The findings are the seven with enough positives to fill every demographic cell. This yields 6{,}048 items from 4{,}054 patients with identical finding and label distributions across demographic groups and 432 positives and 432 negatives per finding. We dropped pneumonia because only 5 positive cases were available for males aged 80+, and merged the original 60--80 and 80+ bands into 60+.

Because view position can affect performance~\citep{viewposition2026}, we checked its demographic balance. AP prevalence differed by 3.1 percentage points between sexes and 5.8 points across age bands, making view position unlikely to explain the demographic results.

\subsection{External validation dataset}
\label{sec:siim-data}

To address the report-derived labels in ChestX-ray14, we repeat the pneumothorax evaluation on SIIM-ACR Pneumothorax~\citep{siimacr2019}. It contains 12{,}047 $1024\times1024$ chest X-rays with radiologist-drawn pixel-level pneumothorax masks. We verified that all image and mask files were present and that labels matched mask content in 60 sampled cases.

Positives make up 22.2\% of the data, so we sample 432 positives and 432 negatives to match the ChestX-ray14 evaluation. SIIM-ACR lacks age, sex, and view metadata and annotates only pneumothorax, so we use it to replicate the pneumothorax result rather than the demographic or per-finding analyses. The median mask covers 0.8\% of the image. We divide the positives by mask area into three groups of 144 to test whether failure is limited to small lesions. The largest group spans 1.2--7.2\% of image area.

\subsection{Question format and image conditions}
Each image was paired with one yes/no question about a single finding, using the default format for every model: \emph{``Is there evidence of \{finding\} on this chest X-ray? Answer yes or no, then briefly explain your reasoning. Start your reply with either `Yes' or `No'.''} Section~\ref{sec:prompt-battery} reports a separate prompt-variation battery. This format makes the prediction unambiguous while allowing a brief explanation.

We evaluated each item under four image conditions. The \emph{real} condition uses the original X-ray, resized to $512\times512$ and converted to RGB. The \emph{swap} condition uses a different patient's X-ray with the opposite label for the same finding, matched on AP/PA view, so a model that uses the image should change its answer. The opposite label makes an unchanged answer informative, while view matching avoids an acquisition confound. The \emph{augmented} condition applies small changes that preserve diagnostic content ($\pm2^{\circ}$ rotation, $\pm5\%$ brightness, JPEG quality 85), so a model that uses the image should not change its answer. The \emph{blank} condition uses a uniform gray image with no medical content, and is excluded from the grounding metric because every model answers ``no'' to it. All variants were generated once with a fixed seed and cached so every model received identical inputs. The same question format, image conditions, and preprocessing were applied to SIIM-ACR.

\subsection{Models and inference}
We evaluated five open medical VLMs from 2B to 7B parameters: MedGemma-4B-it (Gemma~3 with a medical SigLIP encoder, gated under HAI-DEF terms), Lingshu-7B (a medical fine-tune of Qwen2.5-VL), Qwen2.5-VL-7B-Instruct (Apache-2.0, the untuned general-purpose base), HuatuoGPT-Vision-7B (also a medical fine-tune of Qwen2.5-VL), and MedVLM-R1 (2B, Qwen2-VL trained with reinforcement learning). Because Lingshu-7B and HuatuoGPT-Vision-7B are both medical fine-tunes of Qwen2.5-VL-7B, comparing them against the untuned Qwen2.5-VL-7B isolates the effect of medical fine-tuning with architecture and pretraining held fixed, so their differences reflect fine-tuning rather than backbone architecture.

All models ran under vLLM~\citep{kwon2023vllm} at temperature 0 with a fixed seed and a 200-token cap (400 for MedVLM-R1, which is designed to emit longer traces). We use deterministic decoding so that any change in answer reflects the image, not sampling noise.

\subsection{Metrics}

We report recall, specificity, and balanced accuracy (their mean) for each finding, using balanced accuracy because always answering ``no'' gives perfect specificity but zero recall, and constant ``yes'' or ``no'' predictions both score 0.500.

We also measure whether answers depend on the image, comparing each real-image answer with those after an opposite-label patient swap and a mild augmentation. A grounded model should flip after the swap but stay stable after augmentation:
\[
\text{grounding}
=
P(\text{flip}\mid\text{swap})
-
P(\text{flip}\mid\text{augment}).
\]
Higher values indicate greater dependence on diagnostic image content. We compute grounding on ground-truth labels, since conditioning on the model's own predictions would introduce selection bias; where a model never flips, grounding is undefined and we report recall and specificity instead.

\section{Results}

\subsection{Aggregate accuracy hides finding-specific failure}

\begin{figure}[t]
\centering

\newsavebox{\panelbox}
\begin{lrbox}{\panelbox}
\begin{minipage}[t]{0.67\textwidth}
\footnotesize

\begin{tcolorbox}[
    colback=green!4,
    colframe=green!45!black,
    boxrule=0.5pt,
    left=4pt,
    right=4pt,
    top=2pt,
    bottom=2pt,
    arc=2pt,
    before skip=0pt,
    after skip=1.5pt
]
\textbf{MedGemma-4B}
\hfill
\textcolor{green!45!black}{\textbf{``Yes''}\ \checkmark}\\
There is evidence of a pneumothorax. The lung appears to be collapsed, and there
is a visible pleural line.
\end{tcolorbox}

\begin{tcolorbox}[
    colback=green!4,
    colframe=green!45!black,
    boxrule=0.5pt,
    left=4pt,
    right=4pt,
    top=2pt,
    bottom=2pt,
    arc=2pt,
    before skip=0pt,
    after skip=1.5pt
]
\textbf{Lingshu-7B}
\hfill
\textcolor{green!45!black}{\textbf{``Yes''}\ \checkmark}\\
The chest X-ray shows a clear demarcation line, indicative of a pneumothorax.
This line represents the edge of the collapsed lung.
\end{tcolorbox}

\begin{tcolorbox}[
    colback=red!4,
    colframe=red!60!black,
    boxrule=0.5pt,
    left=4pt,
    right=4pt,
    top=2pt,
    bottom=2pt,
    arc=2pt,
    before skip=0pt,
    after skip=1.5pt
]
\textbf{Qwen2.5-VL-7B}
\hfill
\textcolor{red!60!black}{\textbf{``No''}\ $\times$}\\
A pneumothorax would typically appear as a clear space between the lung and the
chest wall, which is not present in this image. The lung fields appear to be
fully inflated.
\end{tcolorbox}

\begin{tcolorbox}[
    colback=red!4,
    colframe=red!60!black,
    boxrule=0.5pt,
    left=4pt,
    right=4pt,
    top=2pt,
    bottom=2pt,
    arc=2pt,
    before skip=0pt,
    after skip=0pt
]
\textbf{HuatuoGPT-7B}
\hfill
\textcolor{red!60!black}{\textbf{``No''}\ $\times$}\\
The chest X-ray does not show evidence of pneumothorax. The lung fields are
well-expanded without any visible air outside the lung boundaries, and the
mediastinum is centrally positioned without shift.
\end{tcolorbox}

\end{minipage}
\end{lrbox}

\noindent
\begin{minipage}[t]{0.30\textwidth}
\vspace{0pt}
\includegraphics[
    width=\linewidth,
    height=\dimexpr\ht\panelbox+\dp\panelbox\relax
]{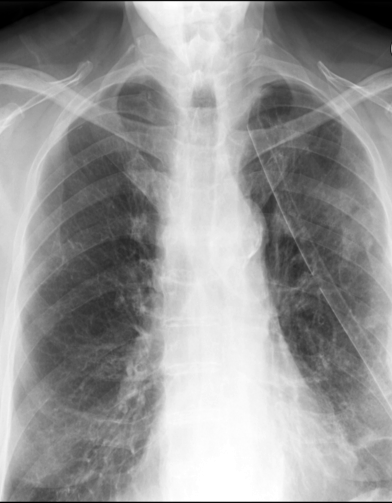}
\end{minipage}%
\hspace{0.008\textwidth}%
\begin{minipage}[t]{0.67\textwidth}
\vspace{0pt}
\usebox{\panelbox}
\end{minipage}

\caption{The same chest X-ray, four models. On a pneumothorax-positive PA film
from ChestX-ray14, MedGemma and Lingshu identify the pneumothorax and describe
the pleural line, while Qwen and HuatuoGPT deny it in fluent, anatomically
specific language, with HuatuoGPT additionally ruling out mediastinal shift.
Across all 432 pneumothorax-positive cases, Qwen and HuatuoGPT return zero
positives.}

\label{fig:silent_failure}
\end{figure}

Table~\ref{tab:permodel} reports aggregate performance across all 6{,}048 items. MedGemma is the most accurate at 0.698, followed by Lingshu at 0.620, while HuatuoGPT (0.534) and Qwen (0.518) are barely above the 0.500 chance level. The yes-rates already hint at the problem, because on a set that is exactly half positive, Qwen answers ``yes'' only 8.9\% of the time and HuatuoGPT 14.7\%. MedGemma and Lingshu depend far more on the image than HuatuoGPT and Qwen, with grounding scores of 0.418 and 0.457 against 0.091 and 0.075, a four to sixfold gap. MedVLM-R1 is degenerate: it answered ``yes'' on only 1.37\% of items (83 of 6{,}048) and scored chance accuracy (0.502), producing none of the expected reasoning traces even under unconstrained prompts, so we exclude it from the per-finding and grounding analyses.

\begin{table}[t]
  \caption{Aggregate performance across 6{,}048 ChestX-ray14 items under the default prompt. Chance accuracy is 0.500. Grounding is conditioned on ground-truth positives and is undefined for MedVLM-R1, which answers ``no'' to 98.6\% of items. Tokens is the mean response length.}
  \label{tab:permodel}
  \centering
  \begin{tabular}{lrrrr}
    \toprule
    Model & Yes rate & Accuracy & Grounding & Tokens \\
    \midrule
    MedGemma-4B    & 0.423 & 0.698 & 0.418 & 37.6 \\
    Lingshu-7B     & 0.488 & 0.620 & 0.457 & 39.9 \\
    HuatuoGPT-7B   & 0.147 & 0.534 & 0.091 & 54.6 \\
    Qwen2.5-VL-7B  & 0.089 & 0.518 & 0.075 & 62.6 \\
    MedVLM-R1      & 0.014 & 0.502 & ---   & \phantom{0}2.0 \\
    \bottomrule
  \end{tabular}
\end{table}

Aggregate accuracy does not show whether performance differs across findings. We therefore examine the positive cases separately in Table~\ref{tab:recall}. Each finding has exactly 432 positive cases. Qwen detects 0 of 432 effusions and 0 of 432 pneumothoraces. HuatuoGPT detects 0 pneumothoraces, 3 effusions, 3 masses, and 1 nodule. However, neither model fails everywhere. HuatuoGPT detects 64\% of infiltration cases, while Qwen detects 40\% of infiltration and 29\% of cardiomegaly cases. The failures are therefore finding-specific. The clearest example is pneumothorax diagnosis, where HuatuoGPT and Qwen detect none of the 432 positive cases while MedGemma detects 38\% and Lingshu 67\% of the same images (Figure~\ref{fig:silent_failure}).

\begin{table}[t]
  \caption{Per-finding recall on 432 positive cases per finding.}
  \label{tab:recall}
  \centering
  \begin{tabular}{lrrrr}
    \toprule
    Finding & MedGemma & Lingshu & HuatuoGPT & Qwen \\
    \midrule
    Atelectasis   & 0.947 & 0.632 & 0.484 & 0.019 \\
    Cardiomegaly  & 0.738 & 0.512 & 0.123 & 0.292 \\
    Effusion      & 0.787 & 0.701 & 0.007 & 0.000 \\
    Infiltration  & 0.604 & 0.586 & 0.641 & 0.400 \\
    Mass          & 0.461 & 0.616 & 0.007 & 0.030 \\
    Nodule        & 0.433 & 0.539 & 0.002 & 0.012 \\
    Pneumothorax  & 0.377 & 0.671 & 0.000 & 0.000 \\
    \bottomrule
  \end{tabular}
\end{table}

Recall counts only positive cases, so it cannot show whether high recall comes at the cost of false positives. Balanced accuracy, the mean of recall and specificity (chance 0.500), fills this in. Qwen and HuatuoGPT sit at 0.499--0.503 on atelectasis, effusion, mass, nodule, and pneumothorax, essentially no discrimination between positive and negative cases. MedGemma shows the opposite imbalance: its 0.947 recall on atelectasis comes with only 0.317 specificity (balanced accuracy 0.632), so it incorrectly flags roughly 68\% of negatives. Lingshu is the only model with balanced accuracy above 0.590 on all seven findings. Aggregate accuracy therefore hides both the finding-specific collapse in HuatuoGPT and Qwen and the recall--specificity imbalance in MedGemma.

\subsection{The pneumothorax failure and its internal control}
\label{sec:prompt-battery}

One worry is that the prompt itself suppressed positive answers. To check this, we tried four prompt formats, the default prompt, a bare question, a forced one-word answer, and a clinical persona. The forced prompt gives a clean internal control for HuatuoGPT. Keeping that prompt fixed and changing only the finding asked about, HuatuoGPT reaches 79.2\% recall on nodule and 69.4\% on effusion, but only 6.2\% on pneumothorax. The prompt clearly draws out positive answers for other findings. It simply fails to do so for pneumothorax.

For nodule and effusion the forced prompt lifts recall, but it lifts false positives by about as much, so balanced accuracy barely changes. On pneumothorax, HuatuoGPT stays near chance under all four prompts, between 0.500 and 0.512.

Qwen is the clearer case. Its pneumothorax recall is exactly 0.000 under every prompt. That is 1{,}728 positive images, seen across a forced one-word answer and a radiologist persona, without a single ``yes''. Its balanced accuracy is 0.500 every time. So HuatuoGPT and the untuned Qwen2.5-VL both miss pneumothorax under every prompt we tried. Lingshu, a medical fine-tune of the same Qwen2.5-VL base, instead detects 67\% of the same cases under the default prompt. The failure is therefore not a quirk of one training run, and on a fixed backbone medical fine-tuning neither reliably causes nor removes it.

The free-text answers show the same thing. HuatuoGPT often writes that the ``lung fields are well-expanded'', even on positive cases, and Qwen describes the sign it should be looking for and then says it is absent. Neither model refuses or expresses any doubt. In a sample of 10 positive cases, all 10 began with ``No''.

\subsection{Ruling out alternative explanations}
\label{sec:alternatives}
We tested the main alternative explanations for the zero pneumothorax recall. Downsizing might erase the thin pleural line, but Qwen detects 0 of 100 positives at both $512$px and native $1024$px. Multi-label images might be harder, but among the 210 pneumothorax-only images Qwen and HuatuoGPT still detect none, while MedGemma and Lingshu detect 35.2\% and 60.0\%. The models might simply answer ``no'' regardless of input, but on the same 100 X-rays both identify chest X-rays and human ribcages at 100\% and reject that the image is mostly black. The failure is therefore specific to diagnosis, not a general inability to use the image or produce positive answers.

One dataset limitation remains: some ChestX-ray14 pneumothorax films are post-treatment, with the pneumothorax drained but a chest tube still visible~\citep{oakdenrayner2020exploring}, and on 150 positives MedGemma and Qwen report a drain in 54.0\% and 52.7\% of cases. This affects absolute recall but not the between-model gap, since on the same images MedGemma still detects 35\% while Qwen detects none. The SIIM-ACR replication (\S\ref{sec:siim-results}) removes the concern with radiologist masks.

A further objection is that these failures are specific to small models, so we evaluated two larger models on all 6{,}048 items. Increasing Qwen2.5-VL from 7B to 72B does not recover pneumothorax: recall rises only from 0.000 to 0.044 (19 of 432) and balanced accuracy from 0.500 to 0.516, still chance, even though the 72B model is not broken, with overall balanced accuracy (0.560) above the 7B's (0.518) and normal detection of other findings such as cardiomegaly (0.631). MedGemma behaves differently, with pneumothorax recall improving from 0.377 at 4B to 0.565 at 27B. Scale can therefore help, but only for some models, and it does not rescue the failing line. Qwen2.5-VL-72B was run as an AWQ 4-bit quantization; because it detects other findings normally, quantization does not explain a failure specific to pneumothorax.

\subsection{External validation on radiologist-annotated data}
\label{sec:siim-results}

Because ChestX-ray14 labels come from radiology reports and its pneumothorax positives often carry a chest drain (\S\ref{sec:alternatives}), we repeat the evaluation on SIIM-ACR, where pneumothoraces have radiologist-drawn pixel-level masks, using the same protocol. The main result replicates (Table~\ref{tab:siim}). HuatuoGPT, Qwen, and MedVLM-R1 detect zero of 432 confirmed pneumothoraces, including zero of the 144 largest lesions, and score 0.500 balanced accuracy. On the same images, Lingshu and MedGemma reach 0.861 and 0.470 recall, and MedGemma's recall rises with lesion size from 0.396 to 0.549. The robustness checks replicate too: Qwen has zero recall under all four prompts, HuatuoGPT stays near chance (0.499--0.517), and both again detect zero of 100 positives at native $1024$px.

\begin{table}[t]
  \caption{SIIM-ACR results on 432 positive and 432 negative images. Positives are grouped by lesion area into terciles of 144; the largest spans 1.2--7.2\% of the image. Grounding is undefined for models that always answer ``no''.}
  \label{tab:siim}
  \centering
  \begin{tabular}{lrrrrrrr}
    \toprule
    Model & Recall & FP & Bal.\ acc. & Small & Mid & Large & Grounding \\
    \midrule
    Lingshu-7B     & 0.861 & 0.250 & 0.806 & 0.847 & 0.903 & 0.833 & 0.613 \\
    MedGemma-4B    & 0.470 & 0.049 & 0.711 & 0.396 & 0.465 & 0.549 & 0.396 \\
    HuatuoGPT-7B   & 0.000 & 0.000 & 0.500 & 0.000 & 0.000 & 0.000 & --- \\
    Qwen2.5-VL-7B  & 0.000 & 0.000 & 0.500 & 0.000 & 0.000 & 0.000 & --- \\
    MedVLM-R1      & 0.000 & 0.000 & 0.500 & 0.000 & 0.000 & 0.000 & --- \\
    \bottomrule
  \end{tabular}
\end{table}

\subsection{Decision confidence: confidently wrong, not merely unsure}
\label{sec:confidence}

A zero recall could reflect genuine uncertainty just short of ``yes'' or a firm commitment to ``no''. To tell these apart, we read the decision from the first answer token, taking the probability on the ``Yes'' and ``No'' tokens (summed over casing variants within the top-20 logprobs), and form two numbers (Table~\ref{tab:confidence}): the normalized confidence $P(\mathrm{Yes})=P_{\mathrm{Yes}}/(P_{\mathrm{Yes}}+P_{\mathrm{No}})$, which is $0.5$ at the decision boundary and near $0$ for a firm ``no'', and the margin $m=\log P_{\mathrm{Yes}}-\log P_{\mathrm{No}}$, which is negative when the model leans ``no''. On pneumothorax-positive images Qwen places almost no probability on ``Yes'' ($P(\mathrm{Yes})=0.026$), the lowest of all seven findings (below even nodule at $0.134$), and its margin of $-3.91$ is the most negative of any finding. HuatuoGPT shows the same pattern more weakly ($0.253$ and $-1.10$), while Lingshu, the same backbone, leans the correct way ($0.715$ and $+1.41$). The failing models are therefore not hesitating near the boundary; they reject pneumothorax firmly, and more firmly than any other finding.

We can also compare how confident each model is with how often it is right. On the balanced pneumothorax set Qwen is $97.6\%$ confident but correct only $50.0\%$ of the time, no better than a coin flip, with an expected calibration error (ECE) of $0.476$. In short, it is confidently wrong, not unsure. MedGemma is also overconfident ($99.2\%$ confidence at $66.3\%$ accuracy, ECE $0.329$) though it still detects $38\%$ of cases, and only Lingshu is both confident and accurate ($84.1\%$ and $76.7\%$, ECE $0.074$). The decision score also carries signal. Ranking all $864$ cases by the margin gives an area under the ROC curve (AUROC, the probability a random positive is scored above a random negative; $0.5$ uninformative, $1.0$ perfect) of $0.586$ for Qwen, above chance, even though it answers ``yes'' to none of these images. The information separating positives from negatives is therefore present in the score; what fails is the threshold, which always lands on ``no''.

\begin{table}[t]
  \caption{Decision confidence on pneumothorax. The $^{+}$ superscript marks the two columns measured on the 432 positive cases only. Margin-AUROC, mean confidence, accuracy, and ECE use all 432 positive and 432 negative cases. Chance accuracy is 0.500.}
  \label{tab:confidence}
  \centering
  \begin{tabular}{lrrrrrr}
    \toprule
    Model & $P(\mathrm{Yes})^{+}$ & Margin$^{+}$ & Margin-AUROC & Mean conf. & Accuracy & ECE \\
    \midrule
    MedGemma-4B    & 0.368 & $-3.19$ & 0.795 & 0.992 & 0.663 & 0.329 \\
    Lingshu-7B     & 0.715 & $+1.41$ & 0.841 & 0.841 & 0.767 & 0.074 \\
    HuatuoGPT-7B   & 0.253 & $-1.10$ & 0.641 & 0.759 & 0.500 & 0.259 \\
    Qwen2.5-VL-7B  & 0.026 & $-3.91$ & 0.586 & 0.976 & 0.500 & 0.476 \\
    \bottomrule
  \end{tabular}
\end{table}

\subsection{Can the model see it? A linear probe on internal representations}
\label{sec:probe}

The results so far concern what the models \emph{say}. To ask what they internally \emph{represent}, we train a linear probe (logistic regression) on the decision-token hidden state to predict whether pneumothorax is present, with patient-grouped 5-fold cross-validation and AUC ($0.5$ not decodable, $1.0$ perfectly decodable; Table~\ref{tab:probe}). We report the last layer of the VLM, the transformer layer whose hidden state feeds the answer, with patient-clustered bootstrap $95\%$ intervals. Because the three Qwen2.5-VL-based models share one backbone, their probes are directly comparable.

The working and failing models separate cleanly. On ChestX-ray14 the last-layer AUC is $0.786$ (Lingshu) and $0.819$ (MedGemma) against $0.613$ (HuatuoGPT) and $0.545$ (Qwen), with non-overlapping intervals. Choosing the single best layer inflates this: Qwen peaks at $0.634$ in layer 1 but falls to $0.578$ once the layer is chosen inside cross-validation, so we report the last layer throughout. Table~\ref{tab:probe} also gives nodule AUC as a within-model comparison.

Running the same probe at every transformer layer of the VLM, from the early layers near the image to the late layers near the output, locates the failure. In the failing models pneumothorax is most decodable early, at layer 1 for Qwen, and then fades toward the output. In the working models it is strongest in the late layers, around layer 27 for Lingshu and layers 19 to 24 for MedGemma, and it is retained to the end where it can drive the answer.

Because many ChestX-ray14 pneumothorax positives carry a chest drain (\S\ref{sec:alternatives}), a probe might detect the tube rather than the pneumothorax. Three checks address this. First, on drain-free positives only, the AUC falls for every model, for example from $0.786$ to $0.598$ for Lingshu, so on ChestX-ray14 the probe does partly rely on the drain. Second, projecting the drain direction out of the features changes the AUC by at most $0.067$, even though the drain is itself highly decodable at $0.70$ to $0.98$. Third, and most important, on SIIM-ACR the label is a radiologist's pixel mask and cannot be a post-drainage artifact. There the probe holds and strengthens, reaching $0.716$ for Qwen and $0.746$ for HuatuoGPT, and $0.851$ for Lingshu. MedGemma, on its different backbone, reaches $0.891$ on the same set. The two models with about $0\%$ recall therefore encode pneumothorax at $0.72$ to $0.75$ on clean labels. The drain-free ChestX-ray14 subset is small, at $n{=}92$, and model-annotated, so SIIM is the decisive test.

Taken together, the failing models represent pneumothorax above chance, most clearly on the clean SIIM labels, yet still answer ``no'': the finding is present internally but weaker than in the working models and is not carried through to the output.

\begin{table}[t]
  \caption{Linear-probe AUC for pneumothorax on the decision-token hidden state (patient-grouped 5-fold CV, last layer, with $95\%$ patient-clustered bootstrap CIs). AUC $0.500$ means not linearly decodable and $1.000$ means perfectly decodable. ``Drain-free'' restricts ChestX-ray14 positives to those without a visible drain ($n{=}92$). Nodule is shown for within-model comparison. $^{\dagger}$MedGemma uses a different backbone, so its SIIM AUC is not directly comparable to the Qwen-family models.}
  \label{tab:probe}
  \centering
  \small
  \begin{tabular}{lcccc}
    \toprule
    & \multicolumn{3}{c}{Pneumothorax AUC} & Nodule \\
    \cmidrule(lr){2-4}
    Model & ChestX-ray14 & Drain-free & SIIM-ACR & ChestX-ray14 \\
    \midrule
    MedGemma-4B   & 0.819 [.791,.848] & 0.631 & 0.891 [.869,.912]$^{\dagger}$ & 0.684 \\
    Lingshu-7B    & 0.786 [.752,.817] & 0.598 & 0.851 [.826,.876] & 0.619 \\
    HuatuoGPT-7B  & 0.613 [.574,.650] & 0.411 & 0.746 [.713,.778] & 0.543 \\
    Qwen2.5-VL-7B & 0.545 [.507,.583] & 0.421 & 0.716 [.683,.751] & 0.549 \\
    \bottomrule
  \end{tabular}
\end{table}

\section{Limitations}
ChestX-ray14's labels are NLP-derived and noisy~\citep{oakdenrayner2020exploring}. The SIIM-ACR replication (\S\ref{sec:siim-results}) addresses this with radiologist masks, but covers only pneumothorax and lacks demographic metadata, so the per-finding and demographic results rest on ChestX-ray14. We ask binary presence questions rather than free-form reports. The demographic analysis found no sex or age disparity after correction for multiple comparisons~\citep{benjamini1995fdr}, but spans only those two axes. The confidence and probe analyses cover pneumothorax alone, and the probe shows the finding is decodable from the hidden states, not that the model uses it. The visual controls also show the working models are not uniformly reliable. We locate where the internal signal fades but not why, so the cause of the failure remains open.

\section{Discussion and conclusion}
Aggregate accuracy is not a safe summary of a medical VLM's diagnostic ability. In our prevalence-matched audit, HuatuoGPT-Vision and Qwen2.5-VL sit at chance on several findings and detect none of the pneumothorax cases. MedGemma, the most accurate model overall, over-calls atelectasis so heavily that its balanced accuracy there collapses. A single top-line number hides both failures. The pneumothorax failure is also robust. It holds across four prompt formats, at native resolution, and on single-label cases. It survives a tenfold increase in model scale. It replicates on SIIM-ACR, where radiologists drew the masks, and there the failing models still miss every case, even the largest lesions. Lingshu is a medical fine-tune of the same backbone, yet it detects most of these cases. The failure is therefore a property of the specific model, not the architecture.

Reading the models' logits and internal states explains how a fluent negative can be so wrong. The failing models are not hesitating. They reject pneumothorax with high confidence, yet they are no better than chance. They are confidently wrong, not unsure. A linear probe still recovers the finding from their hidden states, and most clearly on the clean SIIM labels. The signal is strongest in early layers and fades by the decision layer. So the finding is represented inside the model but suppressed at the output. The information is there. The decision simply lands on ``No'' every time.

The practical message is simple. Medical VLMs should be judged finding by finding, with recall and specificity on balanced cases, because a healthy average can hide a diagnosis the model never sees, and that may be the one a patient can least afford to miss. A fluent answer is not a competent one. When a model names the pleural line of a collapsed lung and then denies it, a clinician has no way to catch the error, and that is what makes this silent failure so dangerous. Because the finding is already encoded internally, the failure may be addressable at the output rather than requiring better perception. Whether a targeted intervention at the output can recover detection, and whether the same pattern holds across other findings and datasets, remains to be tested.

\bibliographystyle{plainnat}
\bibliography{references}

\newpage
\section*{NeurIPS Paper Checklist}

\begin{enumerate}

\item {\bf Claims}
    \item[] Question: Do the main claims made in the abstract and introduction accurately reflect the paper's contributions and scope?
    \item[] Answer: \answerYes{}
    \item[] Justification: The abstract and Section~1 state the finding-specific-blindness claim, the bounded pneumothorax result, and the secondary demographic null, each matched to the corresponding results in Sections~4--6 and stated as counts on a balanced set.

\item {\bf Limitations}
    \item[] Question: Does the paper discuss the limitations of the work performed by the authors?
    \item[] Answer: \answerYes{}
    \item[] Justification: The Limitations section discusses the reliance on NLP-derived ChestX-ray14 labels, the scope of the SIIM-ACR replication, the binary-question and behavioral-only design, the sex/age-only demographic coverage, and the open question of what causes the pneumothorax failure.

\item {\bf Theory assumptions and proofs}
    \item[] Question: For each theoretical result, does the paper provide the full set of assumptions and a complete (and correct) proof?
    \item[] Answer: \answerNA{}
    \item[] Justification: The paper is an empirical audit and contains no theoretical results requiring proof.

    \item {\bf Experimental result reproducibility}
    \item[] Question: Does the paper fully disclose all the information needed to reproduce the main experimental results of the paper to the extent that it affects the main claims and/or conclusions of the paper (regardless of whether the code and data are provided or not)?
    \item[] Answer: \answerYes{}
    \item[] Justification: Section~3 gives the dataset, exact item-construction grid, verbatim question and prompt formats, the four image conditions with their parameters, the five models with versions, and deterministic inference settings (vLLM 0.27.1, temperature 0, fixed seed, token caps). The dataset is public under CC0 and the models are open weights.

\item {\bf Open access to data and code}
    \item[] Question: Does the paper provide open access to the data and code, with sufficient instructions to faithfully reproduce the main experimental results, as described in supplemental material?
    \item[] Answer: \answerNo{}
    \item[] Justification: We do not release code. The underlying dataset (ChestX-ray14) is publicly available under CC0, and Section~3 specifies the item-construction grid, sampling counts, prompts, image conditions, and deterministic inference settings (vLLM 0.27.1, temperature 0, fixed seed) in enough detail to regenerate the evaluation set and reproduce the results, and all five models are open weights.

\item {\bf Experimental setting/details}
    \item[] Question: Does the paper specify all the training and test details (e.g., data splits, hyperparameters, how they were chosen, type of optimizer) necessary to understand the results?
    \item[] Answer: \answerYes{}
    \item[] Justification: No training is performed; all models are evaluated as released. Section~3 specifies the evaluation set construction, the four image conditions and their parameters, prompt formats, and the deterministic decoding settings; the augmentation strength was chosen from measured pixel differences rather than by eye.

\item {\bf Experiment statistical significance}
    \item[] Question: Does the paper report error bars suitably and correctly defined or other appropriate information about the statistical significance of the experiments?
    \item[] Answer: \answerYes{}
    \item[] Justification: Demographic contrasts report patient-level bootstrap 95\% confidence intervals (2{,}000 draws) with Benjamini--Hochberg correction across all 16 tests, and the paper states the minimum detectable effect (0.051 at 80\% power). Headline per-finding results are reported as exact counts on fixed 432-item samples.

\item {\bf Experiments compute resources}
    \item[] Question: For each experiment, does the paper provide sufficient information on the computer resources (type of compute workers, memory, time of execution) needed to reproduce the experiments?
    \item[] Answer: \answerYes{}
    \item[] Justification: All experiments are inference-only, served with vLLM on NVIDIA A100 and RTX PRO 6000 (96\,GB) GPUs. The ChestX-ray14 experiment comprises five models (2B--7B) on 6{,}048 items under four image conditions (120{,}960 generations), plus the four-prompt battery and the robustness and control runs. A scale check adds two larger models (MedGemma-27B and Qwen2.5-VL-72B, the latter as an AWQ 4-bit quantization) on the same 6{,}048 items, and the SIIM-ACR replication repeats the pneumothorax protocol on 864 images. %

\item {\bf Code of ethics}
    \item[] Question: Does the research conducted in the paper conform, in every respect, with the NeurIPS Code of Ethics \url{https://neurips.cc/public/EthicsGuidelines}?
    \item[] Answer: \answerYes{}
    \item[] Justification: The work uses a public, de-identified, CC0 dataset and openly released models, introduces no new data collection from human subjects, and reports safety-relevant failures of medical models to support safer evaluation.

\item {\bf Broader impacts}
    \item[] Question: Does the paper discuss both potential positive societal impacts and negative societal impacts of the work performed?
    \item[] Answer: \answerYes{}
    \item[] Justification: The paper's central point is a patient-safety concern, that medical VLMs can fail silently on specific findings while producing convincing explanations, and it argues for per-finding balanced evaluation as a mitigation. It also cautions that absolute recall numbers are affected by label noise and should not be read as clinical performance.

\item {\bf Safeguards}
    \item[] Question: Does the paper describe safeguards that have been put in place for responsible release of data or models that have a high risk for misuse (e.g., pre-trained language models, image generators, or scraped datasets)?
    \item[] Answer: \answerNA{}
    \item[] Justification: The paper releases no new dataset or model with misuse potential; it evaluates already-public models on an already-public CC0 dataset, and any released code is evaluation-only.

\item {\bf Licenses for existing assets}
    \item[] Question: Are the creators or original owners of assets (e.g., code, data, models), used in the paper, properly credited and are the license and terms of use explicitly mentioned and properly respected?
    \item[] Answer: \answerYes{}
    \item[] Justification: ChestX-ray14 is cited and used under its CC0 release, and SIIM-ACR Pneumothorax (Kaggle challenge data) is used for research under its challenge terms. Each evaluated model is cited, with licenses noted where relevant: Qwen2.5-VL-7B-Instruct under Apache-2.0; Qwen2.5-VL-72B-Instruct, and its AWQ 4-bit quantization, under the Qwen LICENSE (Tongyi Qianwen); and MedGemma (4B and 27B) under the HAI-DEF terms. vLLM is cited as the serving framework.

\item {\bf New assets}
    \item[] Question: Are new assets introduced in the paper well documented and is the documentation provided alongside the assets?
    \item[] Answer: \answerNA{}
    \item[] Justification: We do not release new assets. The prevalence-matched evaluation set is derived from the public ChestX-ray14 dataset and its construction (the 84-cell grid, sampling counts, and cleaning steps) is fully documented in Section~3 so it can be regenerated, but the item manifest and generation scripts are not released.

\item {\bf Crowdsourcing and research with human subjects}
    \item[] Question: For crowdsourcing experiments and research with human subjects, does the paper include the full text of instructions given to participants and screenshots, if applicable, as well as details about compensation (if any)?
    \item[] Answer: \answerNA{}
    \item[] Justification: The paper involves no crowdsourcing and no human subjects; it uses a public de-identified image dataset and automated model inference.

\item {\bf Institutional review board (IRB) approvals or equivalent for research with human subjects}
    \item[] Question: Does the paper describe potential risks incurred by study participants, whether such risks were disclosed to the subjects, and whether Institutional Review Board (IRB) approvals (or an equivalent approval/review based on the requirements of your country or institution) were obtained?
    \item[] Answer: \answerNA{}
    \item[] Justification: The study uses a publicly released, de-identified dataset and involves no interaction with human subjects, so IRB approval is not applicable.

\item {\bf Declaration of LLM usage}
    \item[] Question: Does the paper describe the usage of LLMs if it is an important, original, or non-standard component of the core methods in this research? Note that if the LLM is used only for writing, editing, or formatting purposes and does \emph{not} impact the core methodology, scientific rigor, or originality of the research, declaration is not required.
    \item[] Answer: \answerNA{}
    \item[] Justification: The vision--language models in this paper are the objects of study, not tools used to develop the method; no LLM is a component of the evaluation methodology, so no declaration is required.

\end{enumerate}

\end{document}